\documentclass[twocolumn,aps,prb,superscriptaddress]{revtex4}
\usepackage{mathrsfs}
\usepackage{multirow}
\usepackage{amsmath,amsfonts,amssymb}
\usepackage{array,booktabs}
\usepackage{graphicx,times,graphics,color,epsfig}
\usepackage{extarrows}

\begin{document}
\title{Full Counting Statistics of Phonon Transport in Disordered Systems}

\author{Chao Zhang}
\affiliation{College of Physics and Optoelectronic Engineering, Shenzhen University, Shenzhen 518060, China}

\author{Fuming Xu}
\email[]{xufuming@szu.edu.cn}
\affiliation{College of Physics and Optoelectronic Engineering, Shenzhen University, Shenzhen 518060, China}

\author{Jian Wang}
\email[]{jianwang@szu.edu.cn}
\affiliation{College of Physics and Optoelectronic Engineering, Shenzhen University, Shenzhen 518060, China}

\begin{abstract}
The coherent potential approximation (CPA) within full counting statistics (FCS) formalism is shown to be a suitable method to investigate average electric conductance, shot noise as well as higher order cumulants in disordered systems. We develop a similar FCS-CPA formalism for phonon transport through disordered systems. As a byproduct, we derive relations among coefficients of different phonon current cumulants. We apply the FCS-CPA method to investigate phonon transport properties of graphene systems in the presence of disorders. For binary disorders as well as Anderson disorders, we calculate up to the $8$-th phonon transmission moments and demonstrate that the numerical results of the FCS-CPA method agree very well with that of the brute force method. The benchmark shows that the FCS-CPA method achieves $20$ times more speedup ratio. Collective features of phonon current cumulants are also revealed.

\end{abstract}


\maketitle

\section{introduction}\label{sec0}

Due to the advancement of nanotechnology, the feature size of electronic devices has reached nanoscale. Power dissipation of nano-devices increases as they become smaller which may limit device size and density. Hence it is important to study how power dissipation especially Joule heat is dissipated in nano-devices. Since phonon transport is one of the primary dissipation mechanisms, investigating phonon transport through nano-devices\cite{FOPphonon1,FOPphonon2} can obtain important information in order to design low power devices. From scientific point of view, phonon quantum transport has shown a number of interesting phenomena such as the quantized thermal conductance\cite{universal}, breakdown of Fourier's law in nanoscale systems\cite{break}, phonon filtering\cite{chenphonon}, and topological phononic crystals\cite{topo1,topo2,topo3}, to name just a few.

Impurities are always present in nano-devices giving rise to disorders for phonon transport, which has been studied for various disordered systems. For topological phononic systems, it was found that the topological edge state is intact in the presence of uncorrelated disorder and is gradually destroyed when the disorder is spatially correlated\cite{Lee}. For one-dimensional harmonic chain with mass disorders, the effect of long-range interaction on phonon transport was examined and found to enhance (reduce) the transmission of high (low) frequency phonon\cite{Wang1}. To avoid huge computational burden of the brute force calculation of disorder average, the non-equilibrium vertex correction (NVC) theory\cite{Ke0,Ke01} within coherent potential approximation (CPA)\cite{cpa1,cpa2} was developed to perform analytic average over phonon transmission coefficient\cite{Wang2}. 
When both mass disorder and force constant disorder are present, it was found that the interplay of two disorders gives rise to an anomalous transparency in low dimensional phononic systems\cite{Ke1}. To release the computational burden, a CPA like theory was developed to deal with phonon transmission coefficient with both mass and force constant disorder present\cite{Ke2}. A dynamic cluster approximation formalism was also developed to investigate density of states (DOS) of periodic binary systems with both diagonal and off-diagonal disorders\cite{Mondal}. So far, the CPA formalism has been used to study DOS and phonon transmission coefficient which corresponds to taking disorder average over one and two Green's functions.

In this work, we develop a CPA formalism based on the full counting statistics (FCS) framework that is capable of calculating disorder averaging over $2n$ Green's functions needed for studying $n$-th cumulant of phonon current in the presence of disorders. The central idea of this formalism is to make a nonlinear transformation so that the problem of disorder averaging $2n$ Green's function is mapped into that of averaging one generalized Green's function with one additional parameter, i.e., the counting field\cite{FCSCPA-Fubin}. After this transformation, it is purely a CPA problem and no NVC and other higher order NVC are needed. To obtain the $n$-th cumulant of phonon current, one performs $n$-th order numerical derivative respect to the counting field. We use this FCS-CPA formalism on graphene systems in the presence of binary disorders and Anderson disorders. The phonon transmission moments are calculated up to $8$-th order using FCS-CPA approach. Compared with the brute force method, very good agreement is found. In addition, relations among various coefficients of phonon current cumulants are derived from the symmetry relation of the cumulant generating function.

The paper is organized as follows. In section II, a brief introduction of CPA method is given and the theoretical formalism of FCS-CPA for phonon transport is derived. Two methods of calculating the $n$-th moment of phonon transmission are given: (1). numerical derivatives with respect to the counting field; (2). solving a series of NVC equations. Finally, relations among various phonon current cumulants are derived. In section III, the FCS-CPA formalism is applied to study phonon transport properties of a honeycomb-lattice system and numerical results are presented and analyzed. A short summary is given in section IV.

\section{THEORETICAL FORMALISM}\label{sec1}

\subsection{Coherent potential approximation}

Before introducing the FCS-CPA method, it's necessary to briefly review the coherent potential approximation (CPA). The CPA method can be summarized as: given a complex matrix $A\in \mathbb{C}^{nm\times nm}$, and a block diagonal random matrix $B$ of the same size with each block defined by some independent random distributions $B_{ii}\sim X^{m\times m}$, the average of the matrix inversion can be approximated by
\begin{gather}
\langle \frac{1}{A-B} \rangle = \frac{1}{A-C}, \label{eq:sec21-cpa}\\
	C_{ij}=\left< B_{ii}\left( I-\left( \frac{1}{A-C} \right) _{ii}\left( B_{ii}-C_{ii} \right) \right) ^{-1} \right> \delta_{ij},
\label{eq:sec21-sc}
\end{gather}
where the matrix $C$ is the solution of the above self-consistent equations. It should be noted that we use $i,j$ to label the matrix block, and two more subscripts are required if we want to retrieve one specific matrix element (e.g., $B_{i\alpha,j\beta}$). Such an indexing convention is similar to the tight-binding hamiltonian where the system contains $n$ sites and each site has several orbits (e.g. s/p/d/f-orbit for electron and x/y/z degrees of freedom for phonon).

To derive the above equations, we define $G=(A-C)^{-1}$ which is called the renormalized Green's function and introduce the $T$ matrix such that $(A-B)^{-1} =G + GTG$. It is easy to find
\begin{gather}
\begin{aligned}
    T&=\left( B-C \right) \sum_{i=0}^{\infty}{[ G\left( B-C \right) ] ^i}\\
    &=\left( B-C \right) \left( I+GT \right)\\
    &=\left( I+TG \right) \left( B-C \right).
\label{eq3}
\end{aligned}
\end{gather}
Obviously, $\langle T \rangle =0$ gives rise to Eq.(\ref{eq:sec21-cpa}) and corresponds to CPA.
To facilitate the derivation, it is convenient to define the following three matrices $B_i, C_i, T_i$ as
\begin{gather}
\left( B_i \right) _{jk}=\begin{cases}
    B_{ii},\quad j=k=i,\\
    0,\quad \text{else},\\
\end{cases} \nonumber \\
\left( C_i \right) _{jk}=\begin{cases}
	C_{ik},\quad j=i,\\
	0,\quad \text{else},\\
\end{cases} \nonumber \\
\left( T_i \right) _{jk}=\begin{cases}
	T_{ik},\quad j=i,\\
	0,\quad \text{else},\\
\end{cases} \nonumber
\end{gather}
such that $\sum_i B_i=B$, $\sum_i C_i=C$, and $\sum_i T_i=T$. It should be noted that all these matrices are of the same size as $A$ and the subscript $i$ here denotes that all rows except the $i$-th block row are set to zero. In general, the matrix $C$ is a full matrix. But we will show below, under the single site approximation (SSA), the matrix $C$ is structurally similar to $B$ so that all off-diagonal blocks of $C$ are zero. A visual representation of these matrices look like the following.
\begin{equation}
B_i\sim \begin{pmatrix}
    \Box&		\Box&		\Box\\
    \Box&		\blacksquare&		\Box\\
    \Box&		\Box&		\Box\\
\end{pmatrix} ;\quad C_i ~ {\rm or} ~ T_i\sim \begin{pmatrix}
    \Box&		\Box&		\Box\\
    \blacksquare&		\blacksquare&		\blacksquare\\
    \Box&		\Box&		\Box\\
\end{pmatrix}. \nonumber
\end{equation}
From Eq.(\ref{eq3}) we have
\begin{equation}
T_{ij} = \sum_k (B-C)_{ik} (I + GT)_{kj}, \nonumber
\end{equation}
which is equivalent to
\begin{equation}
T_i=\left( B_i-C_i \right) \left( I+GT_i+G\sum_{j\ne i}{T_j} \right). \nonumber
\end{equation}
Solving for $T_i$, we find
\begin{equation}
T_i=t_i+t_iG\sum_{j\ne i}{T_j}, \nonumber
\end{equation}
where
\begin{gather}
\begin{aligned}
	t_i&=\left( I-\left( B_i-C_i \right) G \right) ^{-1}\left( B_i-C_i \right)\\
	&=\left( B_i-C_i \right) \left( I-G\left( B_i-C_i \right) \right) ^{-1}.\\
\end{aligned}\label{eq:sec21-t-definition}
\end{gather}
So far, no approximation has been made. Now we assume $\langle t_iGT_j \rangle \approx \langle t_i \rangle G \langle T_j \rangle$ for $i \neq j$. This is the so called single site approximation (SSA), which is good at weak disorders where multi-scattering can be neglected. Under this approximation, we have
\begin{equation}
\left< T \right> =\sum_i{\left< t_i \right>}+\sum_{i,j\ne i}{\left< t_i \right> G\left< T_j \right>}. \nonumber
\end{equation}
From the above equation, we see that $\langle t_i \rangle=0$ is a sufficient condition for Eq.(\ref{eq:sec21-cpa}). From $\langle t_i \rangle=0$, we obtain a useful identity,
\begin{gather}
\begin{aligned}
    \left< t_i \right>&=\left< \left( I-\left( B_i-C_i \right) G \right) ^{-1}\left( B_i-C_i \right) \right> \\
    &=\left< \left( I-\left( B_i-C_i \right) G \right) ^{-1}-I \right> G^{-1}, \label{eq:t}
\end{aligned}
\end{gather}
or
\begin{equation}
\quad \left< \left( I-\left( B_i-C_i \right) G \right) ^{-1} \right> =I. \label{eq:cpa-I}
\end{equation}
Inserting this identity back to the first line of Eq.(\ref{eq:t}), we arrive at
\begin{equation}
C_i=\left< \left( I-\left( B_i-C_i \right) G \right) ^{-1}B_i \right>. \nonumber
\end{equation}
Since all columns except the $i$-th block of the rightmost matrix $B_i$ are zero, we can deduce that the only nonzero block of $C_i$ is on the diagonal and equivalently the matrix $C$ is block diagonal. So the self-consistent equation for the $i$-th diagonal block of $C$ can be written as
\begin{equation}
C_{ii}=\left< \left( I-\left( B_{ii}-C_{ii} \right) \left( \frac{1}{A-C} \right) _{ii} \right) ^{-1}B_{ii} \right>. \nonumber
\end{equation}
Similarly, if we use the first line of Eq.(\ref{eq:sec21-t-definition}), we will arrive at Eq.(\ref{eq:sec21-sc}).

Eqs.(\ref{eq:sec21-cpa}) and (\ref{eq:sec21-sc}) frequently appear in calculating transport properties of disordered systems. To evaluate the average of a matrix such as the left hand side of Eq.(\ref{eq:sec21-cpa}), the most intuitive way is the brute force (BF) method. The procedure of brute force calculations is as follows: (1). generating thousands of random samples of matrix $B$; (2). calculating matrix inversions; (3). averaging over the whole ensemble. Apparently, the brute force method is extremely time-consuming. On the contrary, in the CPA method, only a reasonable number of inverse operations are required to solve the self-consistent equations. Although SSA has been made in the derivation, we will show in the numerical results that the accuracy of the CPA simulation is comparable to that of the brute force method.

\subsection{Tight-binding Hamiltonian and Green's function formalism}

\begin{figure}[tbp]
\centering
\includegraphics[width=\columnwidth]{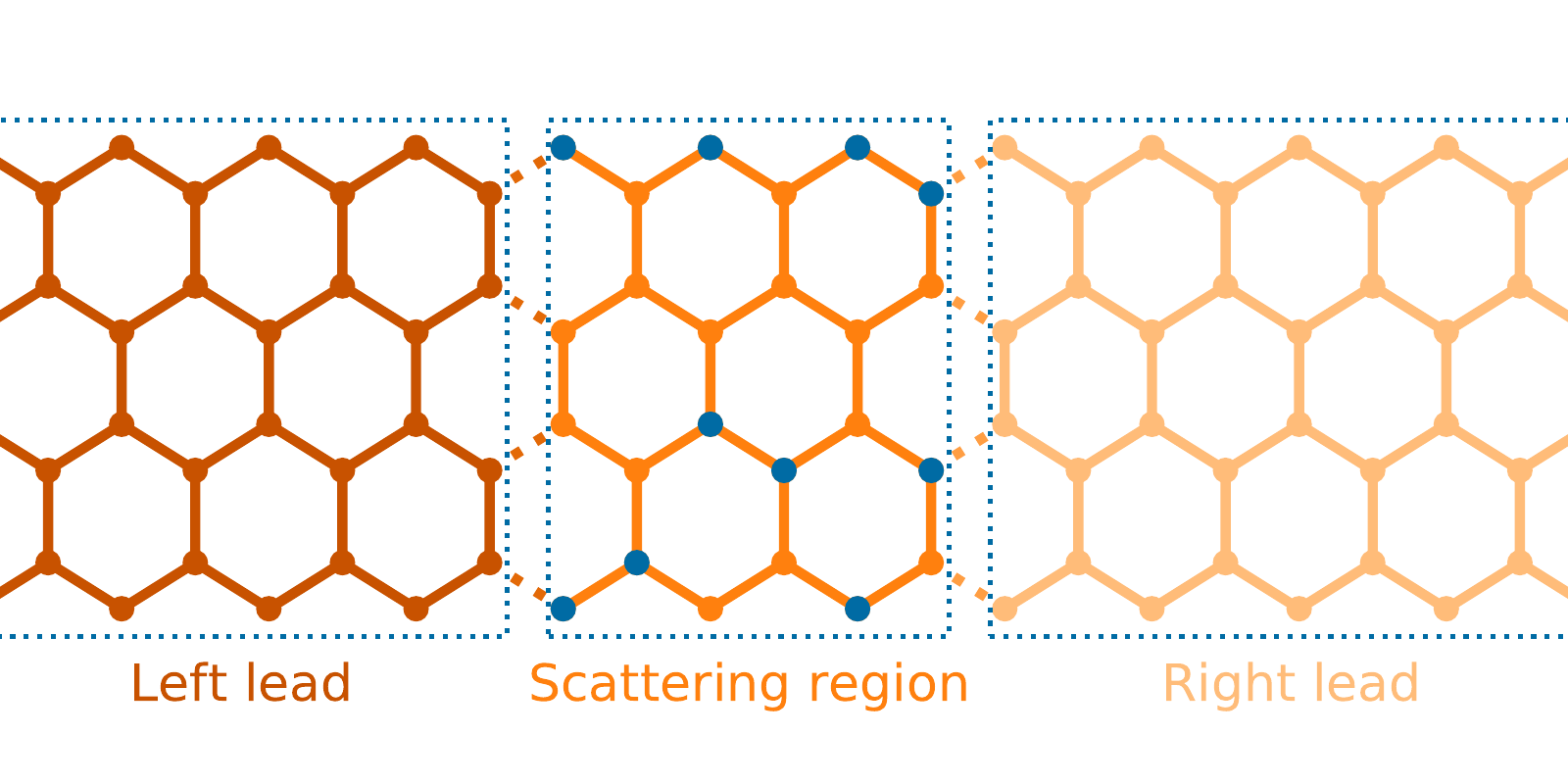}
\caption{Schematic plot of the transport system realized in a graphene nanoribbon with zigzag edges along the transport direction. The system consists of three parts, the central scattering region, the left and right semi-infinite leads. Disordered atoms in the central region are shown in blue color.}
\label{fig:sec22-zigzag-lattice}
\end{figure}

The system we considered is shown in Fig.(\ref{fig:sec22-zigzag-lattice}), which is a zigzag graphene nanoribbon and consists of three parts: the central scattering region, the left and right leads. Driven by the temperature difference between two leads, the phonon flows from the left lead to the right one. For convenience of discussion, we assume that the temperature in the left lead is always higher, $T_L > T_R$. Since both leads are of semi-infinite size, which are also called phonon reservoirs, a steady phonon current is established in the long time limit. For phonon transport, the Hamiltonian is expressed as
\begin{gather}
H=H_C+H_L+H_R+H_T, \nonumber \\
H_C=\sum_{i\in C}{\frac{p_{i}^{2}}{2m_i}}+\sum_{i,j \in C}{\frac{1}{2}K_{ij}x_ix_j}, \nonumber \\
H_L=\sum_{k\in L}{\frac{p_{k}^{2}}{2m_k}}+\sum_{k,k' \in L}{\frac{1}{2}K_{kk^\prime}x_k x_{k^\prime}}, \nonumber \\
H_R=\sum_{k\in R}{\frac{p_{k}^{2}}{2m_k}}+\sum_{k,k'\in R}{\frac{1}{2}K_{kk^\prime}x_k x_{k^\prime}}, \nonumber \\
\begin{aligned}
    H_T=&\frac{1}{2}\sum_{i\in C,k\in L}{\left( K_{ik}x_ix_k+K_{ki}x_k x_i \right)}\\
       +&\frac{1}{2}\sum_{i\in C,k\in R}{\left( K_{ik}x_ix_k+K_{ki}x_k x_i \right)}.
\end{aligned}\nonumber
\end{gather}
The above four terms are Hamiltonians of the central region $H_C$, the left lead $H_L$, the right lead $H_R$, and the coupling $H_T$ between the central region and leads, respectively. The operator $x_i$ describes the displacement from the equilibrium position of the $i$th atom and $p_i$ is the corresponding momentum operator. $x_i$ and $p_j$ satisfy the commutation relation, $[ x_i,p_j ] =i\hbar \delta _{ij}$. The mass term $m_i$ is usually absorbed into the displacement operator $x_i$ for simplification. But this strategy is not available here, since we will treat the mass term as a random variable in disordered systems. We adopt the force constant matrix $K$ from Ref.[\onlinecite{forceConstantModel}] and only consider the first nearest neighbor coupling. The force constant between the neighboring atoms is a $3\times 3$ matrix since each atom has three degrees of freedom (x/y/z), which can be written as
\begin{equation}
K_{ij}=\begin{bmatrix}
    -t_s\cos ^2\theta -t_i\sin ^2\theta&		\left( t_i-t_s \right) \cos \theta \sin \theta&		\\
    \left( t_i-t_s \right) \cos \theta \sin \theta&		-t_s\sin ^2\theta -t_i\cos ^2\theta&		\\
    &		&		t_0\\
\end{bmatrix},\nonumber
\end{equation}
where the stretching factor $t_s$, the in-plane bending factor $t_i$, and the out-of-plane bending factor $t_o$ are obtained by fitting the phonon dispersion relations from experimental data \cite{forceConstantParameter}. To apply the acoustic sum rule, we subtract the summations of those force constants from the diagonal elements as
\begin{equation}
K_{ii}=-\sum_{j\ne i,\alpha =x,y,z}{\mathrm{diag}\left\{ K_{ixj\alpha},K_{iyj\alpha},K_{izj\alpha} \right\}}.
\end{equation}

In terms of the non-equilibrium Greens' function (NEGF), the phonon current is given by the Landauer formula\cite{Landauer1,Landauer2,jwangreview1,chenreview}
\begin{gather}
J_E=\int_0^{\infty}{\frac{d\omega}{2\pi} {\rm Tr}[ \hat{T}]\left( f_L-f_R \right) \hbar \omega}, \label{eq:sec22-phonon-current}
\end{gather}
where ${\hat T}$ is the transmission matrix given by
\begin{gather}
{\hat T}=\Gamma _LG^r\Gamma _RG^a, \nonumber
\end{gather}
and other quantities are defined as
\begin{gather}
G^r =\lim_{\eta \rightarrow 0^+} \frac{I}{M\left( w+i\eta \right) ^2-K_C-\Sigma _{L}^{r}-\Sigma _{R}^{r}},\\ \label{eq:sec22-GreenR}
G^a =\left( G^r \right) ^{\dagger}, \nonumber \\
M =\mathrm{diag}\{ m_1,m_1,m_1,\cdots ,m_N,m_N,m_N \}. \nonumber
\end{gather}

In Eq.(\ref{eq:sec22-phonon-current}),$f_{L/R}=[e^{\hbar \omega/k_B T_{L/R}}-1]^{-1}$ is the Bose-Einstein distribution of the left/right lead, where $k_B$ is the Boltzmann constant, $\hbar$ is the Plank constant, and $\omega$ is the angular frequency of phonon modes. $K_C$ is the force constant matrix of the central region. $M$ is a diagonal matrix and its element $m_i$ represents the mass of the $i$-th atom. The same three $m_i$s in $M$ corresponds to three space degrees of freedoms of each atom. For an open system, the retarded (advanced) Green's function $G^r$ ($G^a$) for the central scattering region can be obtained by absorbing leads' contribution via the self-energy ($\Sigma_L^r$ and $\Sigma_R^r$) and these self-energies can be numerically calculated using the recursive Green's function algorithm\cite{RGF}. Also, the linewidth function $\Gamma_{L/R}=i(\Sigma^r_{L/R}-\Sigma^a_{L/R})$ gives the information about the propagating wavefunction in the lead. The phonon current can be easily understood as follows. Firstly, phonons are propagating from one lead to the other side and only the phonon population difference between two leads $f_L-f_R$ makes the net contribution to the phonon current. Secondly, the transmission coefficient ${\rm Tr}(\hat {T})$ describes the propagation probability for the phonon with a specific energy, and the larger ${\rm Tr}(\hat {T})$ is, the larger the phonon current will be. Lastly, the phonon in all frequency domains will contribute to the current, therefore the integration with respect to the frequency gives rise to the total phonon current.

For phonon transport in disordered systems, we assume that the disorder only exists in the central scattering region. A commonly used disorder type in phonon systems is the isotopic disorder, where the mass of each atom becomes a random variable. If two different isotopes $C_1$ and $C_2$ of the same element exist in the system, the mass of the $i$-th atom $m_i$ is random and has the following distribution,
\begin{equation}
p(m_i) =\begin{cases}
    p,\quad m_i=C_1,\\
    1-p,\quad m_i=C_2.\\
\end{cases} \label{binarydisorder}
\end{equation}
Therefore, the mass matrix $M$ becomes a diagonal random matrix.

In the following, we will mainly focus on the phonon transmission coefficient ${\rm Tr}(\hat {T})$ since it contains all disorder effects in the system. Theoretically, the exact average transmission coefficient $\langle {\rm Tr}(\hat {T}) \rangle$ can be evaluated by enumerating all possible configurations,
\begin{equation}
\langle {\rm Tr}(\hat {T}) \rangle =\sum_{m_1,m_2,\cdots ,m_N}{{\rm Tr}(\hat {T})( m_1,m_2,\cdots ,m_N ) \prod_i p(m_i)}.
\label{eq:sec22-T-average1}
\end{equation}
Obviously, such a method is infeasible for large systems, since it requires $2^N$ samples for a $N$-atom system with binary isotopic diorders. A more practical way is the Monte Carlo method as shown in Eq.(\ref{eq:sec22-T-average}), which we refer as the brute force (BF) method throughout the paper,
\begin{equation}
\langle {\rm Tr}(\hat {T}) \rangle \approx \frac{1}{K}\sum_{i=1}^K{ {\rm Tr}(\hat {T}) \left( {\rm sample_i} \{ m_1,m_2,\cdots ,m_N \} \right)}.
\label{eq:sec22-T-average}
\end{equation}
Here $K$ is the number of random configurations which is far more less than $2^N$. Different from the exact one, the brute force method only takes average over reasonable amount of random samples and still gives a quite accurate result.

\subsection{Phonon transmission moments}

In the previous subsection, we introduced average phonon transport properties of disordered systems. Since the phonon current along is not enough to fully characterize phonon transport properties, we need higher order cumulants or moments of the phonon current operator and the average value is simply the first order cumulant. Full counting statistics (FCS), which has been widely used in transport study such as energy current\cite{YuFSC1, GTangFSC2} and transient dynamics\cite{GTangFSC1,GTangFSC3}, is suitable for evaluating high-order cumulants and moments. Different from the previous work\cite{FCSCPA-Fubin} which focused on the transmission coefficient cumulants, we will show that the transmission moments will give more precise results.

We start with the cumulant generating function\cite{CGFPhonon} (CGF) $\ln \mathcal{Z}$ as shown in Eq.(\ref{eq:sec23-CGF-definition}),
\begin{gather}
\ln \mathcal{Z}=\int_0^{\infty}{\frac{d\omega}{2\pi}\mathcal{X}}, \label{eq:sec23-CGF-definition} \\
\mathcal{X}=-{\rm Tr}\ln (I-\hat {T} Y), \nonumber \\
Y=(e^{i\lambda \hbar \omega}-1) f_L (1+f_R) + (e^{-i\lambda \hbar \omega}-1) f_R (1+f_L). \nonumber
\end{gather}
The $n$-th cumulant of the phonon current $\mathcal{C}_n$ is obtained as the $n$-th derivative of the CGF with respect to $i\lambda$ at $\lambda$=0, \begin{equation}
\mathcal{C}_n=\left. \frac{\partial^n  \ln \mathcal{Z}}{(\partial i\lambda)^n} \right|_{\lambda =0}=\int_0^{\infty}{\frac{d\omega}{2\pi}\left. \frac{\partial^n \mathcal{X} }{(\partial i\lambda) ^n} \right|_{\lambda =0}}. \label{eq35}
\end{equation}
Expanding $\mathcal{X}$ in terms of $\lambda$ and taking the derivative, we find
\begin{gather}
\left. \frac{\partial ^n \mathcal{X} }{(\partial i\lambda)^n} \right|_{\lambda =0}=\sum_{m=1}{\frac{1}{m} {\rm Tr}(\hat {T}^m)  Y_{mn}}, \label{eq:sec23-T-moment-BF}
\end{gather}
where
\begin{gather}
Y_{mn}=\left. \frac{\partial ^nY^m}{( \partial i\lambda ) ^n} \right|_{\lambda =0}. \nonumber
\end{gather}

Eq.(\ref{eq:sec23-T-moment-BF}) can be used to perform brute force calculations of $\langle \mathcal{C}_n \rangle$ where $Y_{mn}$ does not depend on disorders. We now formulate the FCS-CPA method and use it to calculate $\langle \mathcal{C}_n \rangle$ or equivalently $\langle {\rm Tr}(\hat {T}^m) \rangle$. From Eq.(\ref{eq:sec23-T-moment-BF}) we see that the calculation of $\langle \mathcal{C}_n \rangle$ amounts to calculate disorder average of multi-Green's functions. The key advantage of FCS-CPA is to transform the representation of multi-Green's functions to the representation of single generalized Green's function so that CPA can be directly applied\cite{LZhangCPA}. Specifically, one can transform the CGF so that it contains only one generalized Green's function. We first use the following relation with the fact that $\ln \rm {Det} G = \rm {Tr} \ln G$
\begin{equation}
\mathcal{X} = \ln \rm {Det}( I + {\tilde G}~ \Gamma_0 ~ \alpha) = \rm {Tr} \ln ( I + {\tilde G}~ \Gamma_0 ~ \alpha), \label{eq38}
\end{equation}
where $\alpha = \sqrt {Y}$ and
\begin{gather}
\Gamma_0 =\left[ \begin{smallmatrix} & \Gamma_R\\ \Gamma_L& \\ \end{smallmatrix} \right] ~, ~ {\tilde G}=\left[ \begin{smallmatrix} G^r& \\ & G^a \\ \end{smallmatrix} \right]. \nonumber
\end{gather}
Writing Eq.(\ref{eq38}) in an integral form, we have
\begin{gather}
\begin{aligned}
    \mathcal{X}&=-{\rm Tr} \int_0^{\alpha}{dx\frac{1}{\Gamma_0 x+{\tilde G}^{-1}}\Gamma_0}\\
    &=-{\rm Tr} \int_0^{\alpha}{dx\frac{1}{\Gamma_0 x+{ G}^{-1}-B}\Gamma_0},
\end{aligned}\nonumber
\end{gather}
where
\begin{gather}
G^{-1}=-\left[ \begin{smallmatrix} K_C+\Sigma_L^r + \Sigma_R^r& \\ & K_C+\Sigma_L^a+\Sigma_R^a \\ \end{smallmatrix} \right], \nonumber \\
B=-\left[ \begin{smallmatrix} M(\omega+i\eta)^2& \\ & M(\omega-i\eta) ^2\\ \end{smallmatrix} \right]. \label{eq:sec23-B-definition}
\end{gather}
The denominator has been divided into two parts, one is a constant matrix $(\Gamma_0x+G^{-1})$ and the other is a random diagonal matrix $B$. Then we can apply the CPA to obtain the average quantity
\begin{gather}
\langle \mathcal{X} \rangle = -{\rm Tr}\int_0^{\alpha}{dx\frac{1}{\Gamma_0 x+G^{-1}-\Delta(x)}\Gamma_0}, \label{eq:sec23-X-CPA} \\
\begin{aligned}
    \Delta_{ii}(x) &= \Bigl\langle  B_{ii}\biggl( I-\\
    &\left( \frac{1}{\Gamma_0 x+G^{-1}-\Delta \left( x \right)} \right) _{ii}\left( B_{ii}-\Delta_{ii}(x) \right) \biggr) ^{-1} \Bigr\rangle.
\end{aligned} \label{eq:sec23-CPA-self-consistent}
\end{gather}
We note that $\Delta_{ii}(x)$ is a $6 \times 6$ matrix not $3 \times 3$, since the random matrix $M$ is duplicated in Eq.(\ref{eq:sec23-B-definition}). In order to obtain various cumulants, we need to take derivatives of Eq.(\ref{eq:sec23-X-CPA}). The first derivative of $\langle \mathcal{X} \rangle$ with respect to $i\lambda$ gives
\begin{gather}
\frac{\partial \langle \mathcal{X} \rangle}{\partial i\lambda} = -\frac{\partial \alpha}{\partial i\lambda} {\rm Tr} \left[ \frac{1}{\Gamma_0 \alpha +G^{-1}-\Delta(\alpha)}\Gamma_0 \right]. \label{eq45}
\end{gather}
Apparently, the integration is avoided. Therefore, a direct way of calculating the $n$-th cumulant of phonon current is by taking $(n-1)$-th derivative on Eq.(\ref{eq45}) numerically with small enough $\lambda$. Alternatively, we can apply the Taylor expansion on Eq.(\ref{eq:sec23-CPA-self-consistent}) and derive the following equations
\begin{gather}
\Delta(x) =\Delta _0+\sum_{k=1}^{\infty}{\Delta _kx^k}, \label{eq:sec23-Delta-Taylor}\\
\frac{1}{\Gamma_0 x+G^{-1}-\Delta \left( x \right)}=N_0+\sum_{k=1}^{\infty}{N_kx^k}, \label{eq:taylor-Nk}\\
\begin{aligned}
    \Bigl( I-\Bigl( \frac{1}{\Gamma_0 x+G^{-1}-\Delta \left( x \right)} \Bigr) _{ii}& \left( B_{ii}-\Delta _{ii}\left( x \right) \right) \Bigr) ^{-1}=\\
    \left( K_0 \right) _{ii}&+\sum_{k=1}^{\infty}{\left( K_k \right) _{ii}x^k}.
\end{aligned} \label{eq:sec23-K-Taylor}
\end{gather}
Here matrices $N_k$ and $K_k$ are Taylor expansion coefficients and detailed expressions are shown in the Appendix A. Inserting Eq.(\ref{eq:sec23-Delta-Taylor}) and Eq.(\ref{eq:sec23-K-Taylor}) into Eq.(\ref{eq:sec23-CPA-self-consistent}), we can get
\begin{gather}
(\Delta_k)_{ii}=\langle B_{ii}\left( K_k \right) _{ii} \rangle . \label{eq:sec23-FCSCPA-self-consistent-k}
\end{gather}
When $k=0$, Eq.(\ref{eq:sec23-FCSCPA-self-consistent-k}) can be decoupled into two self-consistent CPA equations of block size 3 and they are conjugate to each other, which means only one self-consistent CPA equation is needed to solve. Such a property can be easily derived from Eq.(\ref{eq:sec23-CPA-self-consistent}) by setting $x$=0. When $k=1$, Eq.(\ref{eq:sec23-FCSCPA-self-consistent-k}) actually corresponds to the NVC equation\cite{FCSCPA-Fubin} which is necessary to calculate the average transmission coefficient. For $k=2$, Eq.(\ref{eq:sec23-FCSCPA-self-consistent-k}) is solved to find the average second cumulant of the phonon current when $\Delta_0$ and $\Delta_1$ are provided. Hence for $k>1$, Eq.(\ref{eq:sec23-FCSCPA-self-consistent-k}) will be referred as the higher order NVC equation. Once $\Delta_0,\Delta_1,\cdots,\Delta_{k-1}$ have been established, Eq.(\ref{eq:sec23-FCSCPA-self-consistent-k}) becomes linear with respect to $\Delta_k$ (analysis see Appendix A) and only one more matrix inversion is required to solve for $\Delta_k$. With these Taylor expansion coefficients, we can carry out the derivatives on Eq.(\ref{eq:sec23-X-CPA}) with respect to $i\lambda$
\begin{equation}
\begin{aligned}
    \left. \frac{\partial ^n\left< \mathcal{X} \right>}{\left( \partial i\lambda \right) ^n} \right|_{\lambda =0}=&-\left. \frac{\partial ^n}{\left( \partial i\lambda \right) ^n} \right|_{\lambda =0}\left( \sum_{k=0}^{\infty}{\frac{1}{k+1}{\rm Tr}\left( N_k\Gamma_0 \right) \alpha  ^{k+1}} \right) \\
    =&-\sum_{m=1}{\frac{1}{2m}{\rm Tr}\left( N_{2m-1}\Gamma_0 \right) Y_{mn}}.
\end{aligned}
\end{equation}
In the second line, we drop the even terms of $N_{2m}$ since the trace of its product with $\Gamma_0$ is equal to zero. Compared with Eq.(\ref{eq:sec23-T-moment-BF}), we arrive at
\begin{equation}
\langle {\rm Tr} ( \hat {T}^m ) \rangle =-{\rm Tr}\left( N_{2m-1}\Gamma_0 \right) /2. \label{eq:sec23-T-moment-FCSCPA}
\end{equation}
Eq.(\ref{eq:sec23-T-moment-FCSCPA}) provides us two ways to calculate phonon transmission moments. The left hand side of Eq.(\ref{eq:sec23-T-moment-FCSCPA}) is for the brute force method via averaging over thousands of random samples, and the right hand side corresponds to the FCS-CPA method which is efficient and time-saving.

\subsection{Relations among coefficients of phonon current cumulants}

In the previous subsection, we have introduced phonon current cumulants defined as the derivative of the cumulant generating function (CGF) with respect to $i\lambda$. In terms of $v = 1/k_B T_L - 1/k_B T_R$, we define the coefficient of phonon current cumulants as
\begin{equation}
C_n = \left. \frac{\partial ^n \mathcal{X} }{(\partial i\lambda)^n} \right|_{\lambda =0}=\sum_{m=0}^{\infty} \frac{v^m}{m!} C_n^{(m)}. \label{eq53}
\end{equation}
In this subsection, we will examine the relations among various coefficients $ C_n^{(m)}$. We start from the symmetry relation of the CGF
\begin{equation}
\mathcal{X}[-i\lambda,v] = \mathcal{X}[i\lambda+v,v]. \label{eq54}
\end{equation}
Taking the derivative of Eq.(\ref{eq54}) with respect to $v$, we have
\begin{equation}
\frac{\partial \mathcal{X}[-i\lambda,v]}{\partial v} = \frac{\partial \mathcal{X}[i\lambda+v,v]}{\partial v} + \frac{\partial \mathcal{X}[i\lambda+v,v]}{\partial i\lambda}. \label{eq55}
\end{equation}
From the definition of cumulants and Eq.(\ref{eq53}), we have the Taylor expansion of $\mathcal{X}$ in terms of both $i\lambda$ and $v$,
\begin{equation}
\mathcal{X}=\sum_{n=1}^{\infty}{\sum_{m=0}^{\infty}{\frac{\left( i\lambda \right) ^n}{n!}\frac{v^m}{m!}C_{n}^{\left( m \right)}}}. \label{eq56}
\end{equation}
Inserting Eq.(\ref{eq56}) into Eq.(\ref{eq55}), the left hand side of Eq.(\ref{eq55}) becomes
\begin{equation}
\mathrm{L.H.S.}=\sum_{n=1}^{\infty}{\sum_{m=0}^{\infty}{\frac{\left( i\lambda \right) ^n}{n!}\frac{v^m}{m!}\left( -1 \right) ^nC_{n}^{\left( m+1 \right)}}}. \label{eq57a}
\end{equation}
After rearranging the summation order (see Appendix B), the right hand side of Eq.(\ref{eq55}) becomes
\begin{equation}
\begin{aligned}
    \mathrm{R.H.S.}=&\sum_{n=1}^{\infty}{\sum_{m=0}^{\infty}{\sum_{r=0}^n{\frac{\left( i\lambda \right) ^r}{n!}\binom{n}{r}\frac{v^{m+n-r}}{m!}\left( C_{n}^{\left( m+1 \right)}+C_{n+1}^{\left( m \right)} \right)}}}\\
    =&\sum_{n=1}^{\infty}{\sum_{m=0}^{\infty}{\sum_{s=0}^{m+1}{\frac{\left( i\lambda \right) ^n}{n!}\frac{v^m}{m!}\binom{m+1}{s}C_{n+s}^{\left( m+1-s \right)}}}}.
\end{aligned}\label{eq58}
\end{equation}
Comparing Eq.(\ref{eq57a}) with Eq.(\ref{eq58}), their corresponding coefficients should be equal which gives
\begin{equation}
\left( -1 \right) ^nC_{n}^{\left( m+1 \right)}=\sum_{s=0}^{m+1}{\binom{m+1}{s}C_{n+s}^{\left( m+1-s \right)}},
\end{equation}
where $m=0,1,2,\cdots$.

For $m=0$, we have
\begin{equation}
\begin{cases}
    C_{2n-1}^{\left( 0 \right)}=0,\\
    2C_{2n-1}^{\left( 1 \right)}+C_{2n}^{\left( 0 \right)}=0. \\
\end{cases}\label{eq60}
\end{equation}
From Eq.(\ref{eq60}) we see that $C_{2n-1}$ contains no equilibrium contribution and vanishes when the temperature gradient is zero. In contrast, $C_{2n}$ has both contributions from equilibrium and non-equilibrium processes. The results for $m=1,2,3,4,5$ are listed here,
\begin{align}
m=1: ~~& \quad C_{2n-1}^{\left( 2 \right)}+C_{2n}^{\left( 1 \right)}=0 \label{eq60t}, \\
m=2: ~~& \quad 2C_{2n-1}^{\left( 3 \right)}+3C_{2n}^{\left( 2 \right)}+C_{2n+1}^{\left( 1 \right)}=0, \nonumber \\
m=3: ~~& \quad C_{2n-1}^{\left( 4 \right)}+2C_{2n}^{\left( 3 \right)}+C_{2n+1}^{\left( 2 \right)}=0, \nonumber \\
m=4: ~~& \quad 2C_{2n-1}^{\left( 5 \right)}+5C_{2n}^{\left( 4 \right)}+4C_{2n+1}^{\left( 3 \right)}+C_{2n+2}^{\left( 2 \right)}=0, \nonumber \\
m=5: ~~& \quad C_{2n-1}^{\left( 6 \right)}+3C_{2n}^{\left( 5 \right)}+3C_{2n+1}^{\left( 4 \right)}+C_{2n+2}^{\left( 3 \right)}=0. \nonumber
\end{align}
These five equations can be summarized as
\begin{align}
m=2r-1: ~~& \quad \sum_{i=0}^r{\binom{r}{i}C_{2n-1+i}^{\left( 2r-i \right)}}=0, \nonumber \\
m=2r: ~~& \quad \sum_{i=0}^{r+1}{\begin{Bmatrix} r+1 \\ i \end{Bmatrix} C_{2n-1+i}^{\left( 2r+1-i \right)}}=0, \nonumber
\end{align}
where $\left\{\begin{smallmatrix} r+1 \\ i \end{smallmatrix} \right\}$ are coefficients of $x^i$ in the Taylor expansion of $(2x+1)(x+1)^r$ given by
\begin{equation}
\begin{Bmatrix} r+1 \\ i \end{Bmatrix} =\begin{cases}
    2,\quad i=r+1, \\
    \binom{r}{i}+2\binom{r}{i-1},\quad 0<i<r+1, \\
    1,\quad i=0. \\
\end{cases}
\end{equation}
Using these relations, we can express the coefficients of even cumulants $C_{2n}$ in terms of coefficients of odd cumulants $C_{2n-1}$.

\section{NUMBERICAL RESUTLS}\label{sec2}

\subsection{Phonon transmission moments}

With the developed FCS-CPA method for phonon transport, we numerically study phonon transmission moments and cumulants of the graphene system shown in Fig.\ref{fig:sec22-zigzag-lattice} to demonstrate the accuracy and efficiency of this method. In the calculation, a $4\times 4$ honeycomb lattice has 16 atoms in the central scattering region, while a larger $16\times 16$ lattice hosts 256 sites. The force constant parameters of graphene are adopted from Ref.[\onlinecite{forceConstantParameter}] that the stretching factor $t_s=365.0 J/m^2$, the in-plane bending factor $t_i=245.0 J/m^2$, and the out-of-plane bending factor $t_o=98.2 J/m^2$. The brute force (BF) method, used as a comparison, always take average over $1000$ configurations.

\begin{figure}[tbp]
\centering
\includegraphics[width=\columnwidth]{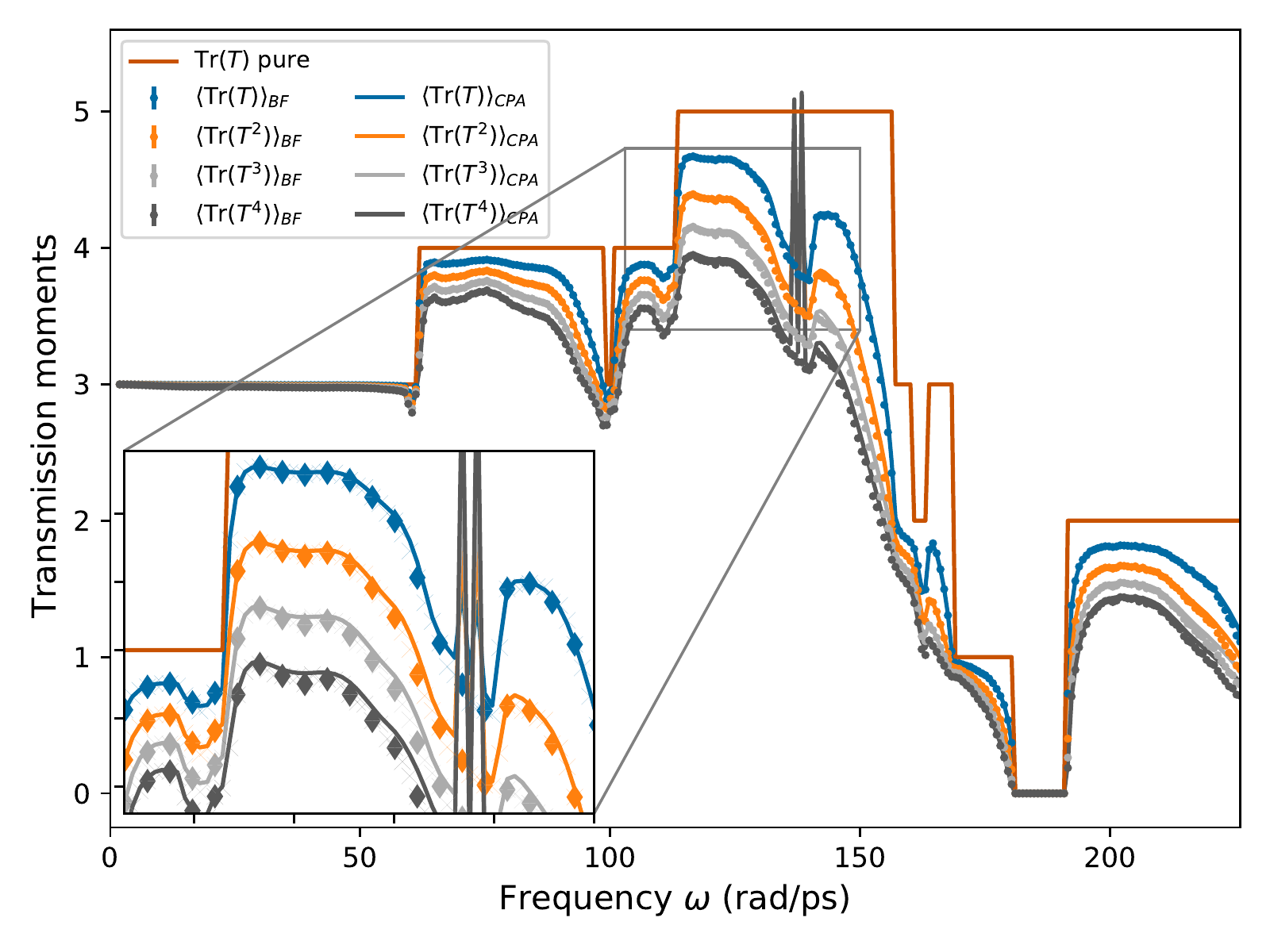}
\caption{Transmission moments as a function of the phonon frequency $\omega$ calculated by BF and FCS-CPA methods for a $4\times 4$ honeycomb lattice with binary disorder. BF results are plotted in dots with error-bar and FCS-CPA results are in solid lines. Transmission spectrum for the pure system is plotted in brown solid line.}
\label{fig:phonon_T_moment4}
\end{figure}

First, we calculate the phonon transmission moments using both BF and FCS-CPA methods for a small $4\times 4$ honeycomb lattice system. The atomic mass in left and right leads is $12$ while the mass in the central region is chosen to be $12$ or $14$ with equal probability to simulate two carbon isotopes, which is called the binary disorder. The first four orders of transmission moments by two methods are presented in Fig.(\ref{fig:phonon_T_moment4}) with FCS-CPA results plotted in solid line while BF results in dots. We use various colors to represent different order of moments. For BF results, the estimated standard deviation is plotted as the error-bar in the figure. The transmission coefficients for the system without disorders are also plotted for reference, which shows clear step behavior. For such a small lattice, its transmission spectrum is quite flat and simple, so we can investigate the disorder influence closely. The small range of error-bar in the figure indicates that the BF method achieve a good convergence. Also, the solid lines go through the dotted data quite well, which suggests that the FCS-CPA results are in good agreement with the BF results in this frequency range. Because the CPA method is only valid in the weak disorder regime while the effective disorder strength in the phonon system $\Delta m\omega^2$ grows rapidly, the FCS-CPA results at higher angular frequency, which is not plotted in the figure, are found to have large error or even diverge. Similar issue is also observed in electron systems. Compared with the quantized transmission in the pure system, all average transmission moments with disorders become much smoother. Transmission moment $\langle {\rm Tr} ({\hat T}^m) \rangle$ decreases when $m$ increases due to the fact that all transmission eigenvalue are less than one, except the near-zero frequency region where transmission moments stay at the value of $3$ which is enforced by the acoustic sum rule.

\begin{figure}[tbp]
\centering
\includegraphics[width=\columnwidth]{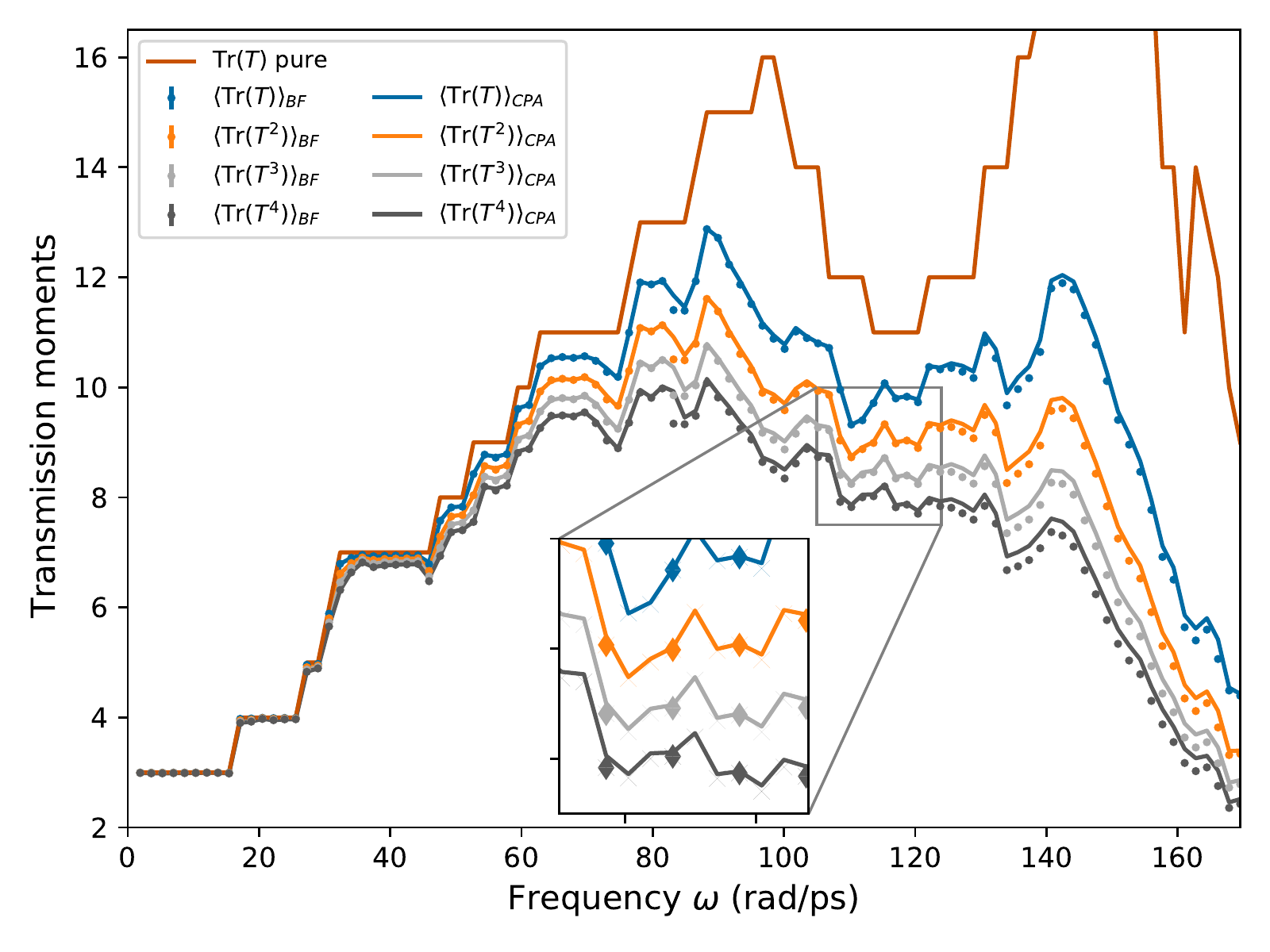}
\caption{Transmission moments versus frequency by BF and FCS-CPA methods for the $16\times 16$ honeycomb lattice with binary disorder. BF results are plotted in dots with error-bar and FCS-CPA results are in solid lines. Transmission spectrum for the pure system is plotted in brown solid line.}
\label{fig:phonon_T_moment16}
\end{figure}

\begin{figure}[tbp]
\centering
\includegraphics[width=\columnwidth]{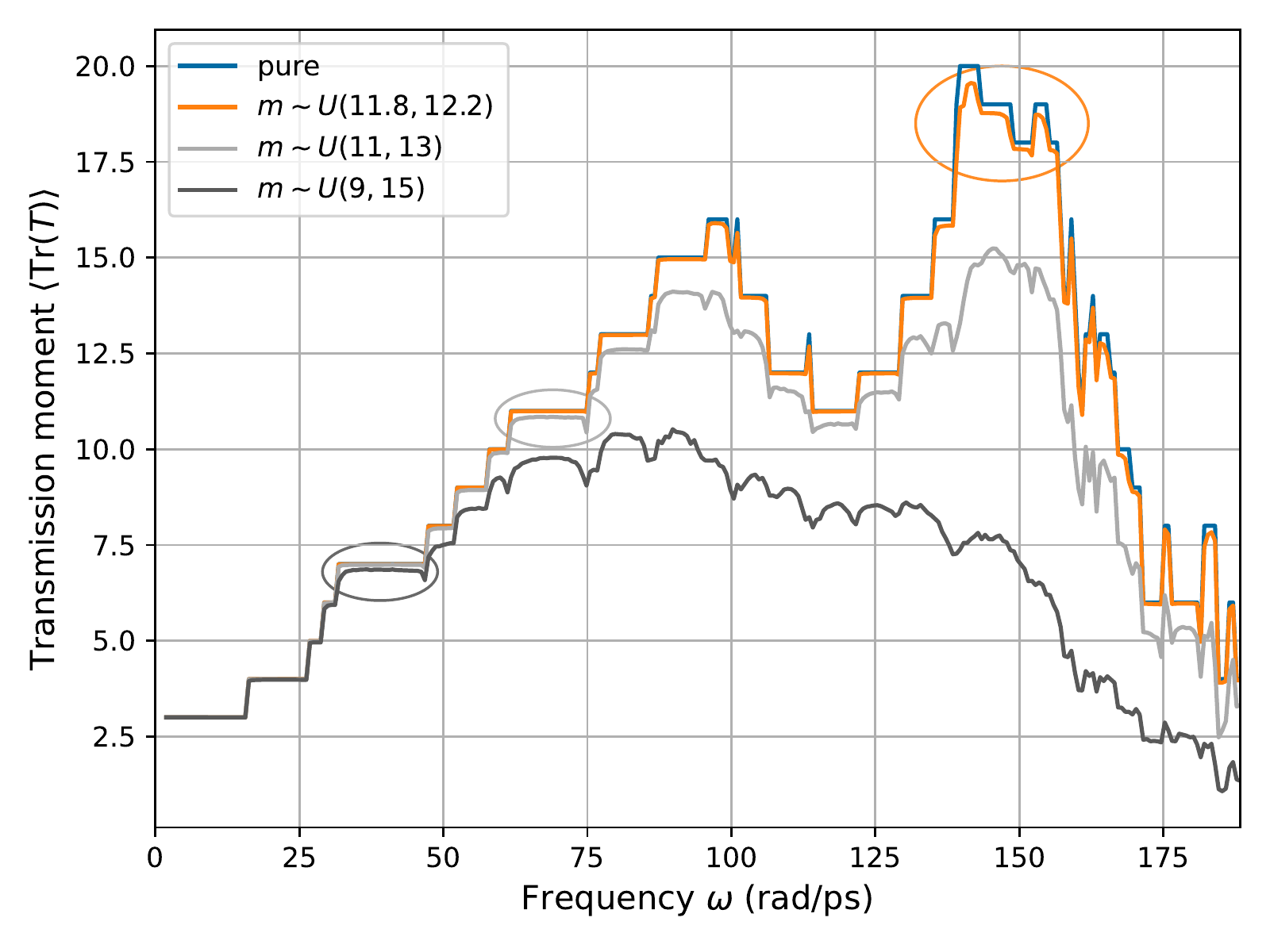}
\caption{First transmission moment of a $16\times 16$ honeycomb lattice with Anderson disorder. Transmission spectrum for the pure system is plotted in blue solid line. The ellipse emphasizes the starting point that the moment clearly deviates from the pure transmission.}
\label{fig:phonon_T_moment_uniform}
\end{figure}

Then, we conduct the same calculation on a binary-disordered $16\times 16$ lattice system and the results are shown in Fig.(\ref{fig:phonon_T_moment16}). Similar to the $4\times 4$ system, the FCS-CPA results are in agreement with the BF results in the wide range of phonon frequency. One of the major differences between the disordered electronic and phononic systems is that the disorder strength for the electron system does not depend on the electron energy, while for the phonon system it is proportional to the square of phonon energy (equivalently the frequency). To demonstrate such difference, we calculate the transmission moments of a $16\times 16$ lattice system with Anderson type disorder. The atomic mass in two leads is $12$ as before while in the central scattering region the mass takes value from a uniform distribution. The first transmission moment $\langle {\rm Tr} ({\hat T}) \rangle$ for different disorder distributions are shown in Fig.(\ref{fig:phonon_T_moment_uniform}). Uniform distributions $U(11.8,12.2)$, $U(11,13)$ and $U(9,15)$ with disorder strength $\Delta m=0.2,1,3$, respectively, are shown in the orange, gray and black lines. Also, the transmission coefficient for the system without disorder is plotted in blue line. The regions where the moments with different disorder strength deviate from the pure transmission spectrum with similar amount are marked with ellipses which are near $\omega=40,70,150$, respectively. For these three regions (or three different disorder strength), the quantity $\Delta m \omega^2=4500,4900,4800$ are close suggesting that the effective disorder strength is indeed proportional to $\Delta m \omega^2$. Numerical results on the second transmission moment $\langle {\rm Tr} ({\hat T}^2) \rangle$ shows similar behaviors.

To benchmark the performance of the FCS-CPA method, we calculate transmission moments of a much larger system. For the phonon transport, we use a $48\times 48$ lattice model and other parameters are kept the same as above. For such a system, the matrix involved in the FCS-CPA calculation is of order $20736$ when solving Eq.(\ref{eq:sec23-FCSCPA-self-consistent-k}). Since each site has only one orbit, a larger lattice $100\times 100$ is used to benchmark for the electronic system and other parameters are adopted from the previous work\cite{FCSCPA-Fubin}. The benchmark is conducted on a workstation of Intel Xeon Gold 5118 CPU with $4$ cores and processor frequency $2.30$ GHz. The results are listed in Table \ref{table:sec312-FCSCPA-benchmark} where the time usage is for one frequency point (energy point for electron). We see that the FCS-CPA method achieves a great speedup ratio, around $20$ times for both phonon and electron systems. The speedup ratio could be even larger if we consider the following aspects:

1) Sampling $1000$ configurations is not accurate enough for the BF method especially for the second and higher order moments and the previous work\cite{FCSCPA-Fubin} suggests that at least $10$ thousands samples should be used.

2) When calculating transmission moments for a series of neighboring frequency points, the solved $\Delta_0$ from the previous phonon frequency could be a good starting point for the next frequency calculation instead of using a random initial value as is done in this benchmark.

\begin{table}[tbp]
\centering
\begin{tabular}{c c c}
    \hline
    & phonon & electron \\
    \hline
    disorder type & binary & uniform \\
    \#orbits per site & $3$ & $1$ \\
    central region size & $48\times 48$ & $100\times 100$ \\
    BF time (s) & $33070$ & $73226$ \\
    FCS-CPA time (s) & $1565$ & $3303$ \\
    speedup ratio & $\times 21.2$ & $\times 22.2$ \\
    \hline
\end{tabular}
\caption{Benchmark of the FCS-CPA method }
\label{table:sec312-FCSCPA-benchmark}
\end{table}

\subsection{Phonon current cumulants}

Once transmission moments are obtained efficiently using the FCS-CPA method, we can evaluate phonon current cumulants by integrating with respect to the frequency through Eq.(\ref{eq35}), and numerical results are shown in this subsection.

\begin{figure}[tbp]
\centering
\includegraphics[width=\columnwidth]{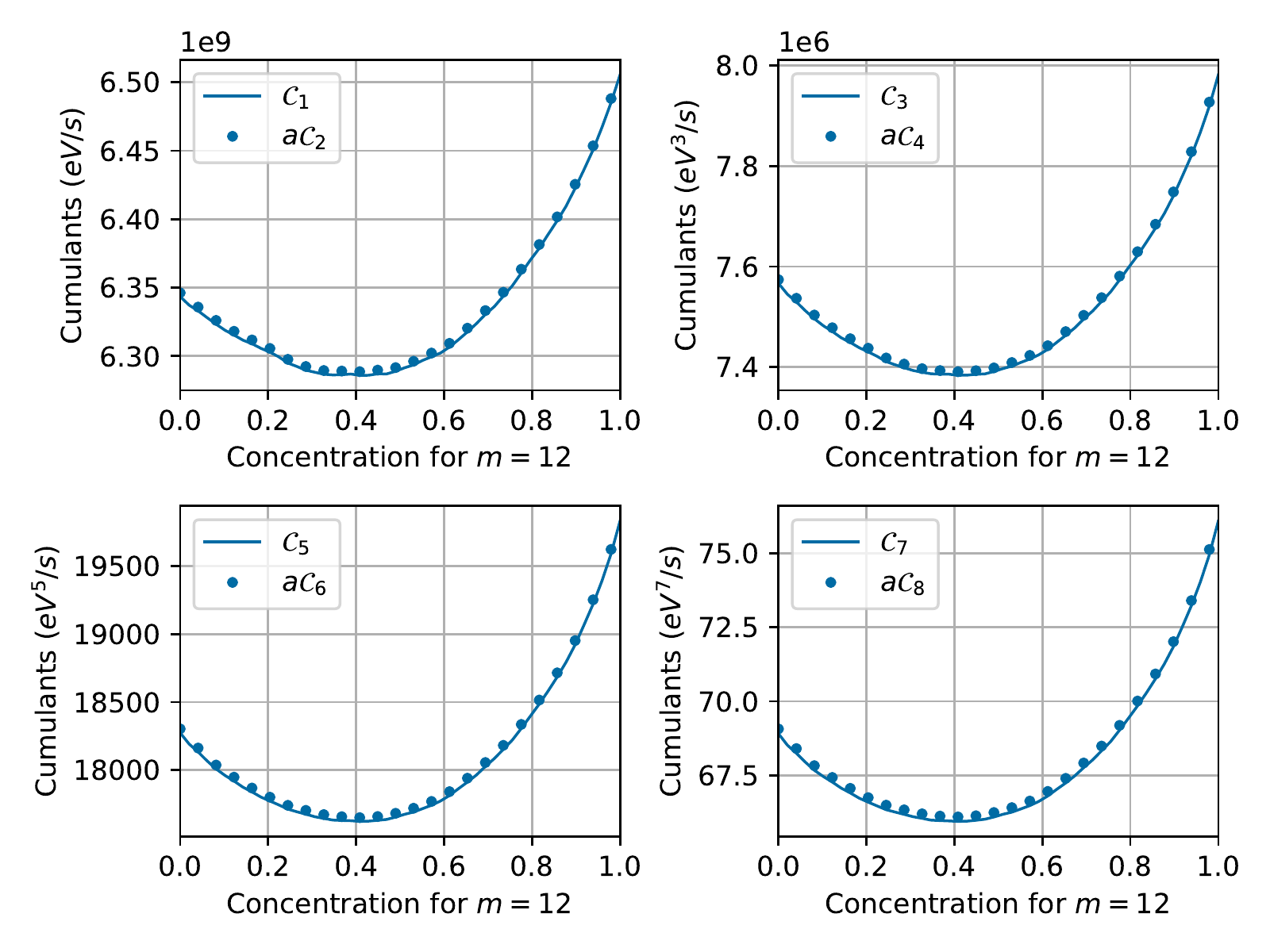}
\caption{Phonon current cumulants of the $16\times 16$ graphene lattice with binary isotopic disorder, versus the concentration of the $m$=12 isotope. The temperatures of two leads are $T_L=101K$ and $T_R=99K$, respectively. All even order cumulants are multiplied by the factor $a=\frac{\Delta T}{2k_B \bar{T}^2}$.}
\label{fig:cumulant_binary_concentration16}
\end{figure}

First, we investigate the influence of binary disorders on phonon current cumulants. The calculation is performed on a $16\times 16$ graphene lattice with binary isotopic disorder. The atomic mass in two leads is set to $m=12$, while in the central region two isotopes with masses $m=12$ and $m=14$ exist with different probabilities, which follows the distribution in Eq.(\ref{binarydisorder}). We set the temperatures in two leads as $T_L=101K$ and $T_R=99K$. Results of the first eight cumulants are shown in Fig.(\ref{fig:cumulant_binary_concentration16}). The probability of the $m=12$ isotope is labeled as the concentration for $m=12$ in the horizontal axis. In the figure, all even cumulants plotted are multiplied by a constant $a=\frac{\Delta T}{2k_B \bar{T}^2}=1.16 eV^{-1}$ which can be deduced from Eq.(\ref{eq60}). Here $\Delta T = T_L - T_R$ is the temperature difference and $\bar{T}=(T_L + T_R)/2$ is the average temperature. As expected in Section II.D., the odd order cumulants are closely matched with the even order cumulants. One can see that, all cumulants decrease as the concentration for $m=12$ increases, reach their minima around concentration $0.4$, and grow rapidly with further increasing of the $m=12$ concentration. Such phenomena also appear in the $12\times 12$ lattice system, suggesting a common behavior.

\begin{figure}[tbp]
\centering
\includegraphics[width=0.95\columnwidth]{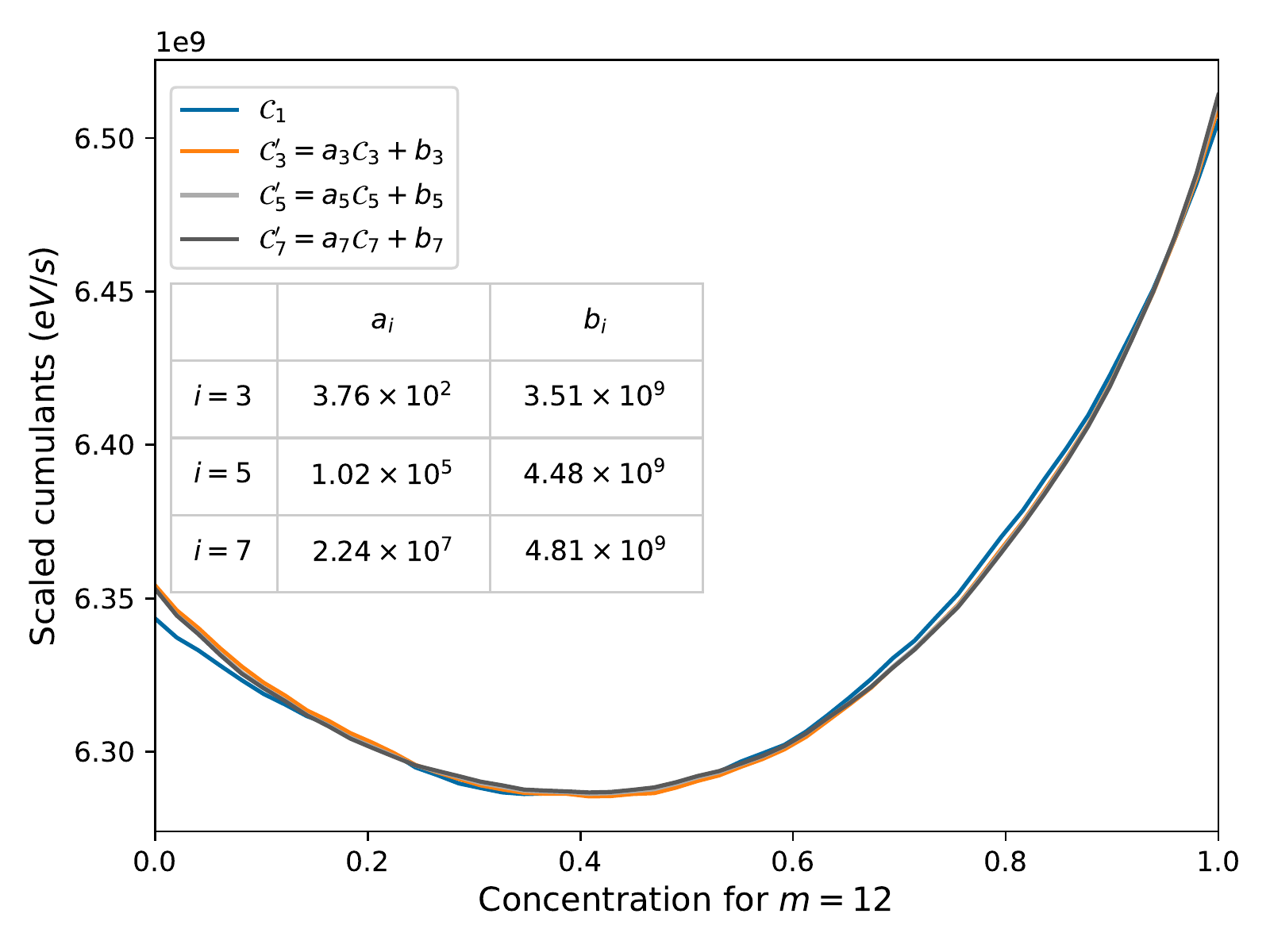}
\caption{Scaled odd cumulants of the $16\times 16$ graphene lattice with binary disorder. The temperatures for two leads are $T_L=101K$ and $T_R=99K$. Higher order cumulants are first normalized to the range $[0,1]$ and then linearly fitted to the first cumulant $\mathcal{C}_1$.}
\label{fig:cumulant_binary_concentration16_4in1_fit0}
\end{figure}

Since all cumulants exhibit similar shapes, we reprocess the odd cumulants in Fig.(\ref{fig:cumulant_binary_concentration16}) and plot the scaled results in Fig.(\ref{fig:cumulant_binary_concentration16_4in1_fit0}). To investigate the underlying connection between these cumulants, we apply the first order polynomial fitting $\mathcal{C}_i' = a_i \mathcal{C}_i + b_i$ to find the best approximation. Higher order cumulants are first normalized to the range $[0,1]$ and then linearly fitted to the first cumulant $\mathcal{C}_1$. The fitting results are shown in Fig.(\ref{fig:cumulant_binary_concentration16_4in1_fit0}) with different colors and coefficients $a_i$ and $b_i$ are displayed in the inset table. Clearly, scaled odd cumulants ${\mathcal{C}_3}'$, ${\mathcal{C}_5}'$, and ${\mathcal{C}_7}'$ are all in good agreement with each other and their corresponding curves almost overlap. They are also close to the first cumulant $\mathcal{C}_1$ and slight deviation appears only in the small concentration region around $[0,0.1]$. Due to the close relation between cumulant coefficients $C_{2n}$ and $C_{2n-1}$, one can expect that even cumulants $\mathcal{C}_4$, $\mathcal{C}_6$, and $\mathcal{C}_8$ have the same linear connection. If we accept the tolerant error in this linear fitting, one can calculate only two points of higher order cumulants with respect to the concentration, and apply curve fitting to obtain other data points. This approximation provides us an efficient way to calculate high order phonon current cumulants.

\begin{figure}[tbp]
\centering
\includegraphics[width=\columnwidth]{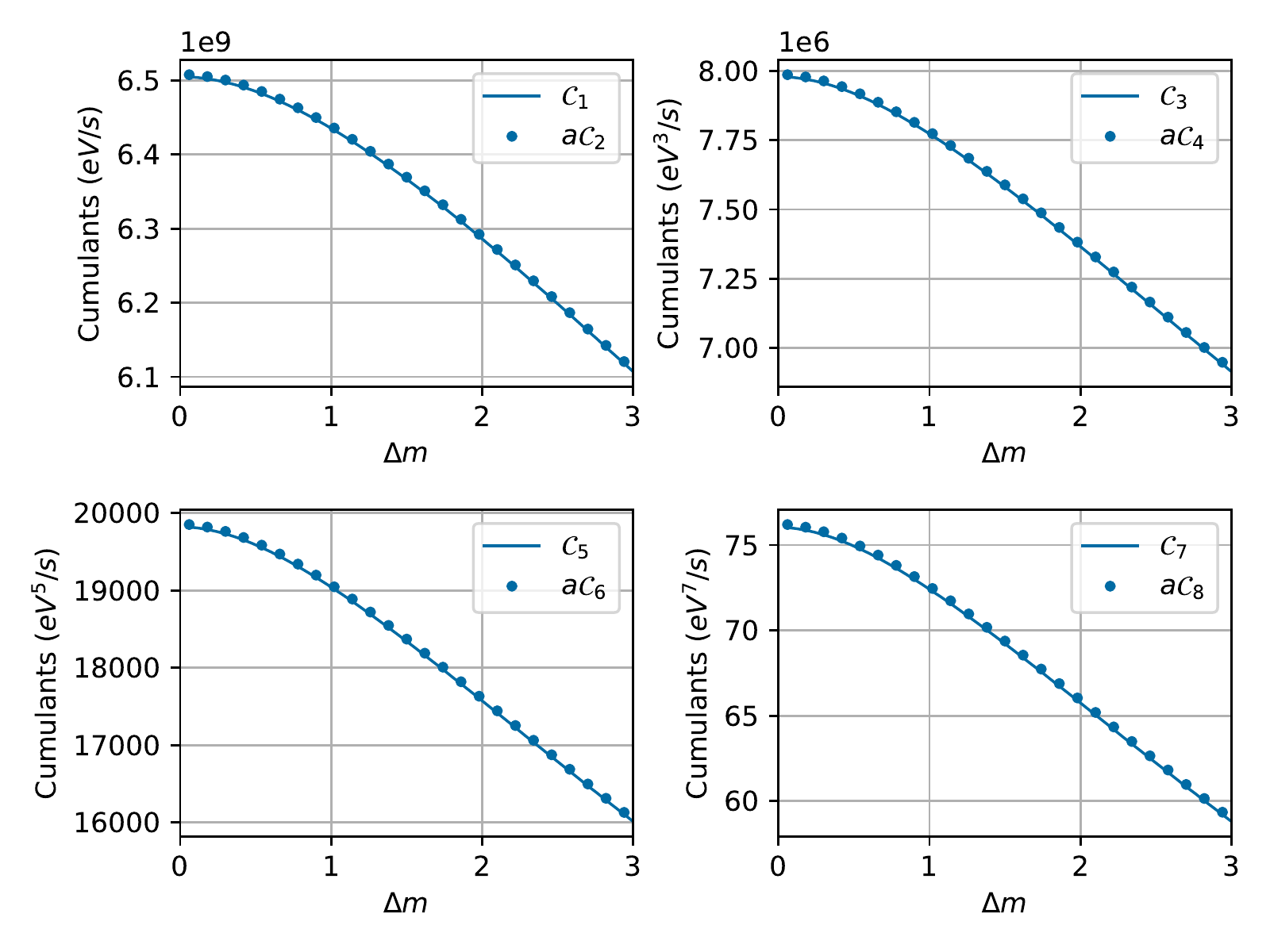}
\caption{Phonon current cumulants of the $16\times 16$ zigzag graphene system with respect to Anderson disorders which uniformly distribute in $U(12-\Delta m, 12+\Delta m)$. The temperatures for two leads are $T_L=101K$ and $T_R=99K$, respectively. All even cumulants are multiplied by the factor $a=\frac{\Delta T}{2k_B \bar{T}^2}$.}
\label{fig:cumulant_uniform_deltaM}
\end{figure}

\begin{figure}[tbp]
\centering
\includegraphics[width=0.95\columnwidth]{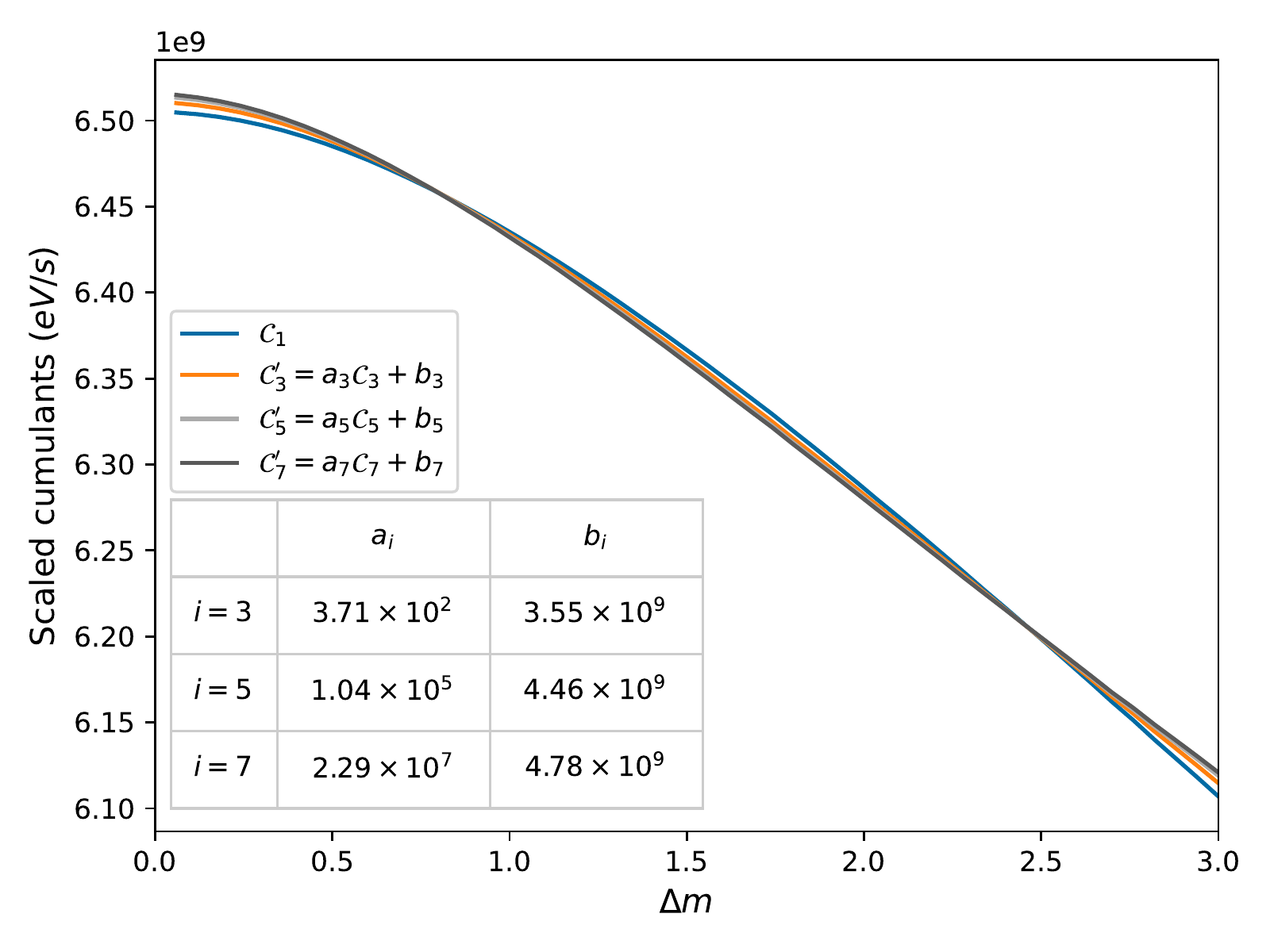}
\caption{Scaled odd cumulants of the $16\times 16$ graphene lattice with Anderson disorder in $U(12-\Delta m, 12+\Delta m)$. The temperatures for two leads are $T_L=101K$ and $T_R=99K$. Higher order cumulants are first normalized to the range $[0,1]$ and then linearly fitted to $\mathcal{C}_1$.}
\label{fig:cumulant_uniform_deltaM_4in1_fit0}
\end{figure}

Then, we study the effect of Anderson disorders on phonon current cumulants. The system under investigation is a $16\times 16$ zigzag graphene nanoribbon with random atomic mass. The randomness follows the Anderson disorder where the mass $m$ is uniformly distributed in $U(12-\Delta m, 12+\Delta m)$ and $\Delta m$ is refereed as the mass disorder strength. Other system parameters are kept unchanged. Cumulants up to the $8$-th order are shown in Fig.(\ref{fig:cumulant_uniform_deltaM}). Different from the binary-disorder case where cumulants show parabolic behavior, all cumulants monotonically decrease with the increasing of $\Delta m$ in the presence of Anderson disorder. Similarly, we see that the odd cumulant $\mathcal{C}_{2n-1}$ matches the even cumulant $\mathcal{C}_{2n}$. Since all cumulants have close shapes, again we use linear fitting $\mathcal{C}_i' = a_i \mathcal{C}_i + b_i$ to reveal their collective features. Odd cumulants from Fig.(\ref{fig:cumulant_uniform_deltaM}) are processed and numerical results are demonstrated in Fig.(\ref{fig:cumulant_uniform_deltaM_4in1_fit0}). One can find that the scaled odd cumulants ${\mathcal{C}_3}'$, ${\mathcal{C}_5}'$, and ${\mathcal{C}_7}'$ are close to $\mathcal{C}_1$ in general, but the similarity is not good as the binary-disorder situation. Small deviations between ${\mathcal{C}_i}'$s exist in the whole disorder strength range, especially at the beginning and ending regions of $\Delta m$. From Fig.(\ref{fig:cumulant_uniform_deltaM}), the even cumulants are expected to behave likely.

\section{Conclusion}

In this work, we developed a theoretical formalism to investigate phononic transport in disordered systems. Our formalism is based on full counting statistics theory combined with the coherent potential approximation. This formalism is capable of calculating disorder averaging of the $2n$-th Green's function, as well as disorder averaging of Green's functions with different energies. We applied our theory to study the full counting statistics of phonon current in the presence of binary or Anderson disorders. The benchmark of our method shows 20 times speedup ratio compared with the brute force method in calculating phonon transmission moments, and the numerical accuracy is kept. Specifically, we have calculated phonon current cumulants up to the $8$-th order. We have also derived a general relation among different coefficients of phonon current cumulants from the symmetry of the generating function, which fits well with numerical results.

\begin{acknowledgments}
This work was financially supported by the National Natural Science Foundation of China (Grant No. 12034014) and the Natural Science Foundation of Guangdong Province (Grant No. 2020A1515011418).
\end{acknowledgments}

\section{appendix}

\subsection{ Taylor expansions of the CPA equation, Eq.(\ref{eq:sec23-CPA-self-consistent}) }

Besides the finite difference method to solve higher order transmission moments as Eq.(\ref{eq45}), we could apply Taylor expansion on Eq.(\ref{eq:sec23-CPA-self-consistent}) to obtain the analytical expression. Since the right hand side of  Eq.(\ref{eq:sec23-CPA-self-consistent}) is quite complicated, we need to conduct the Taylor expansion part by part. First, the part $(\Gamma_0 x+G^{-1}-\Delta (x))^{-1}$ can be expanded as Eq.(\ref{eq:taylor-Nk}) with the following Taylor coefficients
\begin{equation}
N_k=\begin{cases}
    \left( G^{-1}-\Delta _0 \right) ^{-1},\quad k=0, \\
    -N_0\Gamma N_{k-1}+\sum_{m=0}^{k-1}{N_0\Delta _{k-m}N_m},\quad k \neq 0. \nonumber
\end{cases}\label{eq:FCSCPA-taylor-N}
\end{equation}
With those $N_k$, we could expand a larger part as below,
\begin{equation}
\begin{aligned}
    &\left( I-\left( \frac{1}{\Gamma x+G^{-1}-\Delta(x)} \right) _{ii} (B-\Delta(x))_{ii} \right) ^{-1}\\
    =&\left( (H_0)_{ii} + \sum_{k=1}^{\infty}{(H_k)_{ii}x^k } \right) ^{-1}\\
    =&(K_0)_{ii}+\sum_{k=1}^{\infty}{(K_k)_{ii} x^k}, \nonumber
\end{aligned}
\end{equation}
\begin{equation}
(H_k)_{ii}=\begin{cases}
	1-(N_0)_{ii}\left( B-\Delta_0 \right)_{ii},\quad k=0\\
    -(N_k)_{ii}\left( B-\Delta_0 \right)_{ii}\\
    \quad +\sum_{m=1}^k{(N_{k-m})_{ii}(\Delta_m)_{ii}},\quad  k\neq 0,
\end{cases} \nonumber 
\end{equation}
\begin{equation}
(K_k)_{ii}=\begin{cases}
	\left( (H_0)_{ii} \right) ^{-1},\quad k=0, \\
	\sum_{m=0}^{k-1}{(K_0)_{ii}(H_{k-m})_{ii}(K_m)_{ii}},\quad k \neq 0.
\end{cases} \nonumber 
\end{equation}
Here $(N_k)_{ii}$ denotes the $i$-th diagonal block column of the $N_k$ matrix. According to Eq.(\ref{eq:cpa-I}), the matrix $K_k$ should satisfy
\begin{equation}
\langle (K_0)_{ii} \rangle =I,\quad \langle (K_k)_{ii} \rangle =0. \nonumber
\end{equation}
To solve for $\Delta_k$, we need to insert these Taylor expansions into Eq.(\ref{eq:sec23-CPA-self-consistent}) and arrive at the following equation,
\begin{equation}
\begin{aligned}
    &(\Delta_k)_{ii}\\
    = &\langle B_{ii}(K_k)_{ii} \rangle\\
    = &\langle ( B-\Delta_0)_{ii} (K_k)_{ii} \rangle\\
    =& \sum_{j\ne i}{\left< (\Delta_0-B)_{ii}[K_0N_0\Delta_k]_{ij}(N_0)_{ji} (\Delta_0-B)_{ii} (K_0)_{ii} \right>}\\
     +&\sum_{m=1}^{k-1}\Big[{\left< (\Delta_0-B)_{ii} (K_0)_{ii}(H_{k-m})_{ii}(K_m)_{ii} \right>}\\
     +&{\left< (\Delta_0-B)_{ii} (K_0)_{ii}(N_{k-m})_{ii}(\Delta_m)_{ii}(K_0)_{ii} \right>}\\
     +&{\left< (\Delta_0-B)_{ii} (K_0)_{ii}\left[ N_0\Delta _{k-m}N_m \right] _{ii}(\Delta_0-B)_{ii} (K_0)_{ii} \right>}\Big]\\
     -&\left< (\Delta_0-B)_{ii} (K_0)_{ii}\left[ N_0\Gamma N_{k-1} \right] _{ii}(\Delta_0-B)_{ii} (K_0)_{ii} \right>. \nonumber
\end{aligned}
\end{equation}
From above, we can see that only the first term is linear to $\Delta_k$ and other terms are constant matrices once we have $\Delta_0,\Delta_1,\cdots,\Delta_{k-1}$. So $\Delta_k$ can be solved by constructing a linear equations of the form $Ax=b$ where $x$ is a row vector consisting of all elements of $\Delta_k$.

\subsection{Derivation of Eq.(\ref{eq58})}

We start from the first line of Eq.(\ref{eq58}) and change the summation order of $m$ and $r$
\begin{equation}
\begin{aligned}
{\rm Eq.(\ref{eq58})}= &\sum_{n=1}^{\infty}{\sum_{m=0}^{\infty}{\sum_{r=0}^n{\frac{\left( i\lambda \right) ^r}{n!}\binom{n}{r}\frac{v^{m+n-r}}{m!}\left( C_{n}^{\left( m+1 \right)}+C_{n+1}^{\left( m \right)} \right)}}}\\
=&\sum_{r=1}^{\infty}{\sum_{n=r}^{\infty}{\sum_{m=0}^{\infty}{\frac{\left( i\lambda \right) ^r}{n!}\binom{n}{r}\frac{v^{m+n-r}}{m!}\left( C_{n}^{\left( m+1 \right)}+C_{n+1}^{\left( m \right)} \right)}}}. \nonumber
\end{aligned}
\end{equation}

Denoting $k=m+s$ and $s=n-r$, the above expression is equal to
\begin{equation}
\begin{aligned}
=\sum_{r=1}^{\infty}{\sum_{s=0}^{\infty}{\sum_{k=s}^{\infty}{\frac{\left( i\lambda \right) ^r}{\left( s+r \right) !}\binom{s+r}{r}\frac{v^k}{\left( k-s \right) !}\left( C_{s+r}^{\left( k-s+1 \right)}+C_{s+r+1}^{\left( k-s \right)} \right)}}}. \nonumber
\end{aligned}
\end{equation}
Changing the summation order of $s$ and $k$, it goes to
\begin{equation}
\begin{aligned}
=&\sum_{r=1}^{\infty}{\sum_{k=0}^{\infty}{\sum_{s=0}^k{\frac{\left( i\lambda \right) ^r}{\left( s+r \right) !}\binom{s+r}{r}\frac{v^k}{\left( k-s \right) !}\left( C_{s+r}^{\left( k-s+1 \right)}+C_{s+r+1}^{\left( k-s \right)} \right)}}} \\
=&\sum_{r=1}^{\infty}{\sum_{k=0}^{\infty}{\sum_{s=0}^k{\frac{\left( i\lambda \right) ^r}{r!}\frac{v^k}{k!}\binom{k}{s}\left( C_{s+r}^{\left( k-s+1 \right)}+C_{s+r+1}^{\left( k-s \right)} \right)}}} \\
=&\sum_{n=1}^{\infty}{\sum_{m=0}^{\infty}{\sum_{s=0}^m{\frac{\left( i\lambda \right) ^n}{n!}\frac{v^m}{m!}\binom{m}{s}\left( C_{s+n}^{\left( m-s+1 \right)}+C_{s+n+1}^{\left( m-s \right)} \right)}}}, \nonumber
\end{aligned}
\end{equation}
where we have replaced $r$ by $n$ and $k$ by $m$ in the last step. Using the following property of the binomial coefficients,
\begin{equation}
\binom{m}{s}+\binom{m}{s-1}=\binom{m+1}{s}, \nonumber
\end{equation}
we finally arrive at
\begin{equation}
\begin{aligned}
{\rm Eq.(\ref{eq58})}=\sum_{n=1}^{\infty}{\sum_{m=0}^{\infty}{\sum_{s=0}^{m+1}{\frac{\left( i\lambda \right) ^n}{n!}\frac{v^m}{m!}\binom{m+1}{s}C_{n+s}^{\left( m+1-s \right)}}}}. \nonumber
\end{aligned}
\end{equation}


\vspace{2 cm}

\begin{thebibliography}{00}

\bibitem{FOPphonon1}
Z. Liu, X. Yu, K. Chen, Front. Phys. China \textbf{4}, 420 (2009).

\bibitem{FOPphonon2}
R. Su, Z. Yuan, J. Wang, and Z.-G. Zheng, Front. Phys. \textbf{11}, 114401 (2016).

\bibitem{universal} T. Yamamoto, S. Watanabe, and K. Watanabe, Phys. Rev. Lett. \textbf{92}, 075502 (2004).

\bibitem{break} C. Chang, D. Okawa, H. Garcia, A. Majumdar, and A. Zettl, Phys. Rev. Lett. \textbf{101}, 75903 (2008).

\bibitem{chenphonon}
X. Chen, Y. Xu, J. Wang, and H. Guo, Phys. Rev. B \textbf{99}, 064302 (2019).

\bibitem{topo1} P. Wang, L. Lu, and K. Bertoldi, Phys. Rev. Lett. \textbf{115}, 104302 (2015).

\bibitem{topo2} S.H. Mousavi, A.B. Khanikaev, and Z.Wang, Nature Commun. \textbf{6}, 8682 (2015).

\bibitem{topo3} R. Susstrunk and S. D. Huber, Science \textbf{349}, 47 (2015).

\bibitem{Lee} Z.Y. Ong and C.H. Lee, Phys. Rev. B \textbf{94}, 134203 (2016).

\bibitem{Wang1} H.B. Zhou, G. Zhang, J.S. Wang, and Y.W. Zhang, Phys. Rev. E \textbf{94}, 052123 (2016).

\bibitem{Ke0} Y.Q. Ke, K. Xia, and H. Guo, Phys. Rev. Lett. \textbf{100}, 166805 (2008).

\bibitem{Ke01} Y.Q. Ke, K. Xia, and H. Guo, Phys. Rev. Lett. \textbf{105}, 236801 (2010).

\bibitem{cpa1} P. Soven, Phys. Rev. \textbf{156}, 809 (1967).

\bibitem{cpa2} D.W. Taylor, Phys. Rev. \textbf{156}, 1017 (1967).

\bibitem{Wang2} X.X. Ni, M.L. Leek, J.S. Wang, Y.P. Feng, and B.W. Li, Phys. Rev. B \textbf{83}, 045408 (2011).

\bibitem{Ke1} J.X. Zhai, Q.Y. Zhang, Z.H. Cheng, J. Ren, Y.Q. Ke, and B.W. Li, Phys. Rev. B \textbf{99}, 195429 (2019).

\bibitem{Ke2} Z.H. Cheng, J.X. Zhai, Q.Y. Zhang, and Y.Q. Ke, Phys. Rev. B \textbf{99}, 134202 (2019).

\bibitem{Mondal} W.R. Mondal, T. Berlijn, M. Jarrell, and N. S. Vidhyadhiraja, Phys. Rev. B \textbf{99}, 134203 (2019).

\bibitem{FCSCPA-Fubin}
B. Fu, L. Zhang, Y. Wei, and J. Wang, Phys. Rev. B \textbf{96}, 115410 (2017).

\bibitem{forceConstantModel}
J. A. Young and J. U. Koppel, J. Chem. Phys. \textbf{42}, 357 (1965).

\bibitem{forceConstantParameter}
R. A. Jishi, L. Venkataraman, M. S. Dresselhaus, and G. Dresselhaus, Chem. Phys. Lett. \textbf{209}, 77 (1993).

\bibitem{Landauer1}
L. G. C. Rego and G. Kirczenow, Phys. Rev. Lett \textbf{81}, 232 (1998).

\bibitem{Landauer2}
M. P. Blencowe, Phys. Rev. B \textbf{59}, 4992 (1999).

\bibitem{jwangreview1}
J.-S. Wang, J. Wang, and J.T. L\"{u}, Eur. Phys. J. B \textbf{62}, 381 (2008).

\bibitem{chenreview}
X. Chen, Y. Liu, W. Duan, Small Methods \textbf{2}, 1700343 (2018).

\bibitem{RGF}
M. P. Lopez. Sancho, J. M. Lopez. Sancho, J. M. L. Sancho, and J. Rubio, Journal of Physics F: Metal Physics \textbf{15}, 851 (1985).

\bibitem{YuFSC1}
Z. Yu, G.-M. Tang, and J. Wang, Phys. Rev. B \textbf{93}, 195419 (2016).

\bibitem{GTangFSC2}
G.-M. Tang, Z. Yu, and J. Wang, New J. Phys. \textbf{19}, 083007 (2017).

\bibitem{GTangFSC1}
G.-M. Tang and J. Wang, Phys. Rev. B \textbf{90}, 195422 (2014).

\bibitem{GTangFSC3}
G.-M. Tang, Y. Xing, and J. Wang, Phys. Rev. B \textbf{96}, 075417 (2017).

\bibitem{CGFPhonon}
J.-S. Wang, B. K. Agarwalla, H. Li, and J. Thingna, Frontiers of Physics \textbf{9}, 673 (2014).

\bibitem{LZhangCPA}
L. Zhang, B. Fu, B. Wang, Y. Wei, and J. Wang, Phys. Rev. B \textbf{99}, 155406 (2019).


\end{thebibliography}
\end{document}